\documentclass[prd,preprint,showpacs,aps,epsfig]{revtex4}
\usepackage{latexsym}
\usepackage{color}
\usepackage{graphicx}
\usepackage{subfig}
\usepackage{amsmath,amssymb}
\usepackage{makeidx}
\usepackage{hyperref}
\usepackage{subeqnarray}
\usepackage[utf8]{inputenc}
\usepackage{indentfirst}
\usepackage{booktabs}
\usepackage{multirow}
\usepackage{array}
\usepackage{float}
\usepackage{ulem}
\usepackage[papersize={9in,13in}]{geometry}

\graphicspath{{Figuras/}}

\makeindex

\begin{document}

\title{Compact Object with a Local Dark Energy Shell}
	
\author{L. S. M. Veneroni}
\email{leone\_melo@yahoo.com.br}
\author{A. Braz}
\email{antonibraz@gmail.com}
\author{M. F. A. da Silva}
\email{mfasnic@gmail.com}
	
\affiliation{Departamento de F\' {\i}sica Te\' orica, Universidade do Estado do Rio de Janeiro, Rua S\~ ao Francisco Xavier $524$, Maracan\~ a, CEP 20550--013, Rio de Janeiro -- RJ, Brazil}
	
\date{\today}
	
	
\begin{abstract}
	We investigate some models of compact objects in the general relativity theory with cosmological constant $\Lambda$, based on two density profiles, one of them attributed to Stewart and the other one to Durgapal and Bannerji, proposed in the literature to model "neutron stars". For them, a nonlocal equation of state with cosmological constant is obtained as a consequence of the chosen metric. In another direction, we obtain a solution for configurations with null radial  pressure. The first model (based on the Stewart's density profile) turned out to be the most interesting, since surprisingly it admits the presence of dark energy in the interior of the star, in the outermost layers, for a certain range of mass-radius ratio $\gamma$. This dark energy is independent of the cosmological constant, since it is a consequence of  the tangential pressure of the fluid be sufficiently negative. Still in this case, for other values of $\gamma$, all the energy conditions are satisfied. Another advantage of this model, as well as that based on the density profile of Durgapal and Bannnerji is the existence of  intervals of $\gamma$ compatible with physically acceptable models for $\Lambda < 0$, $\Lambda = 0$ and $\Lambda > 0$, which also allowed us to analyze the influence of $\Lambda$ on the behavior of the fluid with respect to the energy conditions. The other configuration studied here, $P_r=0$, only allow solutions for $\Lambda<0$, in order to ensure a positive mass for the object and to satisfy all the energy conditions in a specific range of $\gamma$.
\end{abstract}

	
\maketitle
	
\section{Introduction}

Based on the observational discovery of a recent phase of accelerated expansion of the universe \cite{Riess, Perlmutter}, Einstein's field equations modified by the introduction of a cosmological constant gained a special place in the cosmological scenario, as they could justify a repulsive gravitational effect, as long as it assumes positive values. The additional term involving $\Lambda$ can be interpreted from two points of view: i) Einstein's original equations would not be the best option to describe the theory of gravitation and the introduction of this extra term would correct them, changing the relationship between geometry and matter-energy content; ii) the Universe would be permeated by an unknown energy field and this would be represented by the extra term, which would compose the energy-momentum tensor of the space-time to be described. In the latter case, the parameter $\Lambda$ could be closely associated with the enigmatic dark energy. Considering this last point of view, although the cosmological constant is one of the most promising candidates for dark energy, there is a major problem to be solved, which is to reconcile cosmological data with the interpretation of particle physics, that is, $\Lambda$ would represent a vacuum energy (particle physics), where its scale of energy is enormously greater when compared to the observed energy scale of dark energy (cosmology) \cite{Weinberg}. However, this class of dark energy is not unique. Besides of the dark energy as a modification of the gravitation at large distances, there is another one unrelated to $\Lambda$, which is based on a specific form of matter, such as quintessence, k-essence, and the Chaplygin gas (\cite{Tsujikawa} and references therein). This last class could also appear in a more isolated way in the structure of a compact object for a special fluid, being able to play an important role in the evolution of these objects \cite{BP2005}, \cite{BGS2018}. Like Chaplygin gas, fluids with other equations of state can also act as dark energy, describing exotic fluids with sufficiently negative pressure to violate the strong energy condition, responsible for ensuring the attractiveness of gravitation, that we will refer to hereinafter as local dark energy. Anyway, it is reasonable to question whether the presence of the cosmological constant, with possible important consequences for cosmology, could in some way affect the configuration of a stable or collapsing compact object.
 
Trying to answer this question, several authors have considered the cosmological constant in Einstein's equations. Markovic and Shapiro \cite{MS2000} have studied a spherical homogeneous collapse of a dust cloud. and concluded that a positive cosmological constant  can slow down the collapse initially, although at later times the sphere’s self-gravity pulls the sphere into the final singularity, the last one showed that. Gonçalves \cite{G2001} investigated how a positive cosmological constant could interfere in the singularity formation when it is considered in general inhomogeneous spherical dust collapse. Lake \cite{L2000} have extended the Markovic and Shapiro analyzes to inhomogeneous and degenerate cases. The end state in the collapse of null dust with negative cosmological constant was carried out by Lemos \cite{Lemos}. Deshingkar et al. \cite{DJCJ2001} have analyzed some dust solutions with positive, negative, and null values of the cosmological constant in order to investigate the implications of this towards the final outcome of gravitational collapse. As a remarkable result, they pointed out that the naked singularity formed in the dust collapse can be partly covered by a positive $\Lambda$. These results suggest that the cosmological constant can play an important role in the evolution of compact, collapsing or stable, objects. More recently, Bordbar et al. \cite{BHP2016} have investigated the role of $\Lambda$ in the upper limit for the maximum mass of a stable compact object and showed that, for the considered model, the cosmological constant becomes important only for value greater than $\Lambda^{-14}$s$^{-2}$. However, none of these studies considered the combined effect of the cosmological constant with anisotropy on pressure.

Here we are also interested in the role played by anisotropy in pressure in the description of static objects. Bowers and Liang \cite{BowersLiang} were the first to consider anisotropic for spherically symmetric static configurations. According with Ruderman \cite{Ruderman} the radial pressure may not be equal to the tangential one in massive stellar objects. Anisotropy can be introduced by the existence of a solid core, different kinds of phase transitions, pion condensation or even by combination of different perfect fluid, among others  \cite{MakHarko} and references therein. In a paper published in 2011, Malafarina e Joshi \cite{MalafarinaJoshi} bring up the important discussion on how the pressure only in the form of tangential stresses might direct the collapse towards the formation of black holes or naked singularities. They modified a wellknown black hole formation process, for a dust cloud, introducing an arbitrarily small tangential pressure and  found that it could now go to a naked singularity final configuration. Many others authors have pointed out that the anisotropy can plays a important role on the physical, affecting the critical mass, the stability and the redshift of stars \cite{Herreraetal}-\cite{Hillebrondt}.

In the section that follows this introduction, we present the field equations for a static spherically symmetric matter distribution with anisotropic pressure. Then, we obtain a system of differential equations that directly imply a non-local equation of state for the fluid, already pointed out by Hernández and Núñez \cite {HN2004}, where we now include the cosmological constant. We also presented the physical restrictions, such as the hydrostatic equilibrium, regularity, junction, and energy conditions. In the third section, we applied the results for a particular density profile \cite{Stewart}, while in the section IV we consider other one \cite{DurgapalBannerji}. In addition, we considered the possibility of building models for which the radial pressure is zero, in such a way that we would only have the tangential pressure and the cosmological constant to support the star. The model with the Stewart's profile proved to be particularly interesting, since it admits a range of mass-radius ratio for the star, where only the strong condition is violated, at its edge. In fact, this dark energy, kind a quintessence \cite{Tsujikawa}, which is surprisingly trapped in the edge of the star, has nothing to do with the cosmological constant. Nevertheless, $\Lambda$ may interfere on the models, at least for values much higher than that predicted when it is interpreted as the main cause of the accelerated expansion of the Universe, what is also  discussed. Finally, in the fifth and last section, the main results are summarized and compared with other results in the literature. 

\section{Field equations for the interior spacetime}

The general metric for a spherically symmetric spacetime can be written as
\begin{eqnarray}
\label{met0}
ds^2_{-} &=& g_{\alpha\beta}d\chi^{\alpha}d\chi^{\beta} \nonumber\\
&=& -e^{2 \nu(r)} d t^{2}+e^{2 \lambda(r)} d r^{2}+r^{2}\left[d \theta^{2}+\operatorname\sin^{2}(\theta) d \phi^{2}\right].
\end{eqnarray}

For a spherical distribution of anisotropic fluid at pressures, the momentum energy tensor is given by
\begin{equation}
\label{em}
(T^\alpha_\beta)^{-} =
\left(\begin{array}{cccc}
-\rho & 0 & 0 & 0 \\ 
0 & P_r & 0 & 0 \\ 
0 & 0 & P_\perp & 0 \\ 
0 & 0 & 0 & P_\perp
\end{array}\right) \,.
\end{equation}\

If we consider the cosmological constant, the Einstein's equations are given by
\begin{equation}
\label{eins1}
G^{\alpha}_{\beta}+\Lambda \delta^{\alpha}_{\beta} = R^{\alpha}_{\beta}-\frac{R \delta^{\alpha}_{\beta}}{2}+\Lambda \delta^{\alpha}_{\beta}= 8\pi T^{\alpha}_{\beta} \,,
\end{equation}
with $G=c=1$.

Thus, the nonzero components of the Einstein tensor can be written as
\begin{equation}
\label{neweins1} 
-\frac{1}{r^2}+\frac{e^{-2\lambda}}{r}\left(\frac{1}{r}-2\lambda'\right) + \Lambda = -8\pi\rho \,,
\end{equation} 
\begin{equation}
\label{neweins2}
-\frac{1}{r^2}+\frac{e^{-2\lambda}}{r}\left(\frac{1}{r}+ 2\nu' \right) + \Lambda = 8\pi P_r \,,
\end{equation}
\begin{equation}
\label{neweins3}
e^{-2\lambda}\left[\frac{\nu'}{r}- \frac{\lambda'}{r}+ \nu''- \nu'\lambda'+ (\nu')^2\right] + \Lambda = 8\pi P_\perp \,,
\end{equation}
where prime denotes differentiation with respect to the coordinate $r$.

From (\ref{neweins1}), we obtain
\begin{equation}
\label{neweins1b}
e^{-2\lambda} = 1- \frac{2m(r)}{r}-\frac{\Lambda r^2}{3} \,,
\end{equation} 
where we define $m(r)$ as
\begin{equation}
\label{massa0}
m(r) = 4\pi \int_{0}^{r}\rho(\overline{r})\overline{r}^2d\overline{r} \,.
\end{equation}

\subsection{Hydrostatic equilibrium equation}

Next, we obtain the hydrostatic equilibrium equation for anisotropic fluids with cosmological constant, analogous to TOV, already presented in \cite{BHP2016} for isotropic fluids. For this, taking the derivative of (\ref{neweins2}) and combining the result with (\ref{neweins2}) and (\ref{neweins3}), we get
\begin{equation}
\label{comp5}
8\pi P'_r=\frac{2}{r}\left[8\pi (P_\perp-P_r)\right]-\frac{2e^{-2\lambda}}{r}\left[\lambda'\nu'+(\nu')^{2}\right] \, .
\end{equation}

So, by combining (\ref{neweins1}), (\ref{neweins2}), (\ref{neweins1b}), and (\ref{comp5}), we obtain the Tolman-Oppenheimer-Volkoff (TOV) equation for anisotropic fluids with cosmological constant, that is,
\begin{equation}
\label{tov1}
P'_r=-(\rho +P_r)\left[\frac{12\pi r^{3}P_r +3m-\Lambda r^3}{r\left(3r-6m-\Lambda r^3\right)}\right]+\frac{2}{r}(P_\perp-P_r)\,.
\end{equation}


\subsection{Solutions with a nonlocal state equation}

Let us redefine the two functions $\mu$ and $\nu$ as done by \cite{HN2004} as
\begin{equation}
\label{newvar1}
e^{2\nu} = h(r)e^{4\beta(r)} \,,
\end{equation}
\begin{equation}
\label{newvar2}
e^{2\lambda} = \frac{1}{h(r)} \,,
\end{equation} 
with 
\begin{equation}
\label{newvar3}
h(r) \equiv 1-\frac{2m(r)}{r}-\frac{\Lambda r^2}{3} \,
\end{equation}
and then, rewrite the metric (\ref{met0}) as
\begin{equation}
\label{met0b}
ds^2_{-} = -h(r)e^{4\beta(r)}d t^{2}+ \frac{1}{h(r)}d r^{2}+ r^{2}\left[d \theta^{2}+\operatorname\sin^{2}(\theta) d \phi^{2}\right] \,.
\end{equation}

Hernández, Núñez, and Percoco \cite{HNP1999} showed that the metric above, but dependent on the time, naturally satisfies a nonlocal equation of state. The same is verified in the static limit. This equation of state can be modified by adding a term with a cosmological constant, that is

\begin{equation}
\label{nles}
P_r(r)=\rho(r)-\frac{2}{r^3}\int_{0}^{r}\overline{r}^2\rho (\overline{r})d\overline{r}+\frac{a}{2\pi r^3}+\frac{\Lambda}{6\pi} \,,
\end{equation}
where $a$ is an arbitrary constant.

Einstein tensor components for the metric (\ref{met0b}) can be expressed as
\begin{equation}
\label{neweins4}
\frac{h+h'r-1}{r^2} + \Lambda = -8\pi\rho \,,
\end{equation}
\begin{equation}
\label{neweins5}
\frac{h+h'r-1}{r^2}+ \frac{4h\beta'}{r}+ \Lambda = 8\pi P_r \,,
\end{equation} 
 \begin{equation}
\label{neweins6}
\frac{h'+2h\beta'}{r} + \frac{1}{2}\left[h''+4h\beta''+6h'\beta'+8h(\beta')^2 \right]+ \Lambda = 8\pi P_\perp \,.
\end{equation} 

Differentiating (\ref{nles}) with respect to $r$, we obtain
\begin{equation}
\label{nles2}
P'_r = \rho'+\frac{6}{r^4}\int_{0}^{r}\overline{r}^2\rho(\overline{r})d\overline{r}-\frac{2\rho}{r}-\frac{3a}{2\pi r^4} \,.
\end{equation}

From equation (\ref{nles}), we have
\begin{equation}
\label{nles4}
\frac{6}{r^4}\int_{0}^{r}\overline{r}^2\rho (\overline{r})d\overline{r} = \frac{3\rho}{r}-\frac{3P_r}{r}+\frac{3a}{2\pi r^4}+\frac{\Lambda}{2\pi r} \,.
\end{equation}

Considering (\ref{nles2}), we can rewrite (\ref{nles4}) in term of derivatives as
\begin{equation}
\label{nles6}
\rho- 3P_r+ r(\rho'- P'_r) = -\frac{\Lambda}{2\pi} \,,
\end{equation}

Differentiating (\ref{neweins4}) and (\ref{neweins5}) with respect to $r$, and substituting them into (\ref{nles6}), we get the same expression obtained by \cite{HN2004}, then, the same solution for $\beta(r)$, that is, 
\begin{equation}
\label{beta1}
\beta(r) = \frac{1}{2}\ln\left(\frac{\xi}{h}\right) +\int\frac{a_0 dr}{r^2h} +a_1 \,,
\end{equation}
being $\xi$, $a_0$ and $a_1$ arbitrary constants and, without lost of generality, we can adopt $a_1=0$.

First, using (\ref{newvar3}) in (\ref{neweins4}), we get
\begin{equation}
\label{neweins9}
8\pi\rho = \frac{2m'}{r^2} \,.
\end{equation}

Substituting (\ref{newvar3}) into (\ref{beta1}), and considering (\ref{neweins9}), we find
\begin{equation}
\label{beta4}
\beta(r) = \frac{1}{2}\ln\left(\frac{\xi}{1-\frac{2m(r)}{r}-\frac{\Lambda r^2}{3}}\right) +\int\frac{a_0 dr}{r^2\left(1-\frac{2m(r)}{r}-\frac{\Lambda r^2}{3} \right) } \,.
\end{equation} 

Following, inserting (\ref{newvar3}) and (\ref{beta4}) into (\ref{neweins5}), we have
\begin{equation}
\label{neweins10}
8\pi P_r = \frac{2m'}{r^2}+ \frac{4\Lambda}{3}+ \frac{4(a_0-m)}{r^3} \,.
\end{equation}

Finally, using (\ref{neweins9}) in (\ref{neweins10}), we get
\begin{equation}
\label{massa0b}
m(r) = a_0+2\pi r^3\left(\rho-P_r +\frac{\Lambda}{3} \right) \,.
\end{equation}

Thus, for $\lim_{r\rightarrow0} m(r)=0$, we must have $a_0=0$. So we can denote $\beta$ as 
\begin{equation}
\label{beta5}
\beta = \frac{1}{2}\ln\left(\frac{\xi}{h}\right) \,.
\end{equation}

Therefore, substituting (\ref{beta5}) into (\ref{met0b}), we arrive in the same metric proposed in \cite{HN2004}, that is,
\begin{eqnarray}
\label{met1}
ds^2_{-} &=& g^{-}_{\alpha\beta}d\chi^{\alpha}_{-}d\chi^{\beta}_{-} \nonumber\\
&=& -\frac{\xi^2}{h(r)}dt^{2}+\frac{1}{h(r)}dr^{2}+r^{2}d\theta^{2} + r^{2}\sin^{2}(\theta)d\phi^2 \,,
\end{eqnarray}
and the index ``$-$" refers to the metric that describes the space-time fulfilled by the matter.

\subsection{Junction conditions}
	
As we are considering solutions with cosmological constant $\Lambda$, the exterior of the star should be described by the Schwarzschild-De Sitter metric, that is

\begin{eqnarray}
\label{met2}
ds^2_{+} &=& g^{+}_{\alpha\beta}d\chi^{\alpha}_{+}d\chi^{\beta}_{+} \nonumber\\
&=& -\left(1-\frac{2M}{r}-\frac{\Lambda r^2}{3} \right) dt^{2}+\left(1- \frac{2M}{r}-\frac{\Lambda r^2}{3}\right)^{-1}dr^2+r^2d\Omega^2 \,,
\end{eqnarray}
where $d\Omega^2 =d\theta^2+\sin^{2}(\theta)d\phi^2$.

The interior spacetime is described by the metric (\ref{met1}).
The two regions are separated by a hypersurface, here called $\Sigma$. The metric for this region can be written as
\begin{eqnarray}
\label{met3}
ds^2_{\Sigma} &=& g_{ij}d\varsigma^{i}d\varsigma^{j} \nonumber\\
&=& -d\tau^2+R^{2}(\tau)(d\theta^2+\sin^{2}(\theta) d\phi^2) \,.
\end{eqnarray}
with $\varsigma^{i}=(\varsigma^{1},\varsigma^{2},\varsigma^{3})=(\tau,\theta,\phi)$.

On the hypersurface $\Sigma$, the metrics must match, that is,
\begin{equation}
\label{igual1}
(ds^2_{-})_{\Sigma} \,= \, (ds^{2})_\Sigma \, = \, (ds^2_{+})_{\Sigma} \,.
\end{equation}

Imposing that 
\begin{equation}
\label{xi}
\xi = 1-\frac{2M}{R}-\frac{\Lambda R^2}{3} \,,
\end{equation}\\*
with $r_\Sigma = R$.

Based on (\ref{igual1}), equations (\ref{met1}), (\ref{met2}) and (\ref{met3}) provide
\begin{equation}
\label{igual3}
\frac{\xi^2}{h(R)}\,dt^{2} = d\tau^2 = \left(1-\frac{2M}{R}-\frac{\Lambda R^2}{3} \right) dt^{2} \,,
\end{equation}
where $r = R$ at the hypersurface $\Sigma$.

Thus, from (\ref{xi}) and (\ref{igual3}), we get
\begin{equation}
\label{hR}
h(R) = 1-\frac{2M}{R}-\frac{\Lambda R^2}{3} \,,
\end{equation}
\begin{equation}
\label{igual4}
\left(\frac{dt}{d\tau}\right)^2_\Sigma = \left(1-\frac{2M}{R}-\frac{\Lambda R^2}{3} \right)^{-1} \,.
\end{equation}

The extrinsic curvature to the hypersurface $\Sigma$ is given by
\begin{equation}
\label{cur1}
K^{\pm}_{ij} = -\eta^{\pm}_{\alpha}\frac{\partial^{2}\chi^{\alpha}_{\pm}}{\partial\varsigma^{i}\partial\varsigma^{j}} - \eta^{\pm}_{\alpha}\Gamma^{\alpha}_{\beta\gamma}\frac{\partial \chi^{\beta}_{\pm}}{\partial\varsigma^{i}}\frac{\partial \chi^{\gamma}_{\pm}}{\partial\varsigma^{j}} \,.
\end{equation}

The continuity of the second fundamental form shows us that 
\begin{equation}
\label{cur2}
K^{-}_{ij}|_\Sigma = K^{+}_{ij}|_\Sigma \,.
\end{equation}

The unitary normal vectors to the hypersurface are given by
\begin{equation}
\label{cur12}
\eta^{-}_\alpha= \frac{1}{\sqrt{h(r)}}\left(0,1,0,0\right) \,, 
\end{equation}
\begin{equation}
\label{cur16}
\eta^{+}_\alpha= \sqrt{1-\frac{2M}{r}-\frac{\Lambda r^2}{3}}\left(0,1,0,0\right) \,. 
\end{equation}

The continuity of the component $K_{11}$ of the extrinsic curvature furnishes
\begin{equation}
\label{hdR}
h'(R) = 2\left(\frac{\Lambda R}{3}-\frac{M}{R^2}\right) \,,
\end{equation}
where the continuity of the other components is identically satisfied.

Considering the metric (\ref{met1}), the Einstein's field equations are given by
\begin{equation}
\label{rho}
8\pi\rho(r) = \frac{1-h-rh'}{r^2}-\Lambda \,,
\end{equation}
\begin{equation}
\label{Pr}
8\pi P_r(r) = \frac{h-rh'-1}{r^2}+\Lambda \,,
\end{equation}
\begin{equation}
\label{Pt}
8\pi P_\perp (r) = \frac{(h')^2-hh''}{2h}+\Lambda \,.
\end{equation}

Now, taking (\ref{rho}) and (\ref{Pr}) on the hypersurface $\Sigma$ and considering (\ref{hR}) and (\ref{hdR}), we obtain 
\begin{equation}
\label{PrR0}
P_r(R) = 0 \,
\end{equation}
and
\begin{equation}
\label{rhoR}
\rho(R) = \frac{\Lambda R^2 +6\gamma}{12\pi} \,.
\end{equation}
where $\gamma = M/R$.

\subsection{Energy conditions}

For a solution of the field equations that represents a compact object to be physically reasonable, in addition to satisfying the conditions of regularity and junction with an outer spacetime, it must also satisfy the energy conditions. These are given by \cite{HawkingEllis}

\begin{eqnarray}
\label{energia1}
&&\rho \geq 0 \,,\quad \rho + P_r \geq 0 \,,\quad \rho + P_\perp \geq 0 \,,\quad \rho - P_r \geq 0 \,,\quad \rho - P_\perp \geq 0 \,, \nonumber\\
&&\rho + P_r + 2P_\perp -\frac{\Lambda}{4\pi}\geq 0 \,.
\end{eqnarray}

The last of the inequalities above reveals the influence of both the fluid and the cosmological constant on the attractiveness, for $\Lambda <0$, or repulsiveness of gravitation, for $\Lambda > 0$. For $\Lambda = 0$, this inequality refers only to the fluid, representing one of the conditions strong energy. When the inequality is violated, including the cosmological constant, in this work we adopt the name ``effective strong energy condition"  and  the fluid which violates it is called ``effective dark energy”, in the sense that both the contribution of the fluid and that of the cosmological constant are taken into account for the analysis of the effect of convergence/divergence of a congruence of time-like or null-type geodesics.

In the following, we consider some particular cases as possible stellar models.

\section{A compact object partially constituted by local dark energy}

The Stewart's density profile \cite{Stewart}, which includes the Florides' solution \cite{Florides} as a particular case when the radial pressure is zero, was revisited by Grokhroo and Mehra \cite{GM1994}, H. Hernández and L. A. Núñez \cite{HN2004} and by some of us in a previous paper on gravitational collapse \cite{VS2019}. It is known that this density profile can give rise to an equation of state similar to the Bethe–Börner–Sato \cite{Betheetal} Newtonian equation of state for nuclear matter \cite{Martinez} and references therein. The Stewart's profile can be written as
\begin{equation}
\label{rho1}
\rho(r) = \frac{\sigma}{8\pi}\left(1-\frac{K r^2}{R^2} \right) \, ,
\end{equation}
where $\sigma$ and $K$ are constants to be determined.

From equation (\ref{rho}), we have
\begin{equation}
\label{h1}
h(r) = 1-\frac{\left(\Lambda+\sigma\right)r^2}{3}+\frac{\sigma Kr^4}{5R^2} \,,
\end{equation}
and then, (\ref{Pr}) and (\ref{Pt}) can be put respectively in the forms
\begin{equation}
\label{Pr1}
P_r(r) = \frac{5\sigma R^2-9K\sigma r^2 + 20\Lambda R^2}{120\pi R^2} \,,
\end{equation}
\begin{eqnarray}
\label{Pt1}
P_\perp(r) = \frac{A_0(r)}{120\pi R^2\left(3K\sigma r^4-5R^2(\Lambda +\sigma)r^2+15R^2 \right)} \,.
\end{eqnarray}
where
$A_0(r)=18K^2\sigma^2r^6 +15KR^2\sigma(2\Lambda-\sigma)r^4+25R^4r^2(\sigma^2-\Lambda \sigma-2\Lambda^2) -270KR^2\sigma r^2 +75R^4(\sigma+4\Lambda).$

In the center of the star we have isotropic pressures, as can be seen below,
\begin{equation}
\label{isotropia}
P_r(0)=P_\perp(0)=\frac{\sigma+4\Lambda}{24\pi} \,.
\end{equation}

The junction condition (\ref{PrR0}) couples the parameter $\sigma$ with the cosmological constant, that is
\begin{equation}
\label{sigma}
\sigma = \frac{20\Lambda}{9K-5} \,,
\end{equation} 
since $K\neq 5/9$.

Considering (\ref{h1}) at $r=R$ and comparing with (\ref{hR}), we get
\begin{equation}
\label{K}
K=\frac{5}{3}\left(\frac{2\Lambda R^2+3\gamma}{2\Lambda R^2 +9\gamma} \right) \,,
\end{equation}

Substituting (\ref{K}) into (\ref{sigma}), we find
\begin{equation}
\label{sigma1}
\sigma = \frac{2\Lambda R^2+9\gamma}{R^2} \,,
\end{equation}
which allows us to write
\begin{equation}
\label{rho2}
\rho(\delta) = \frac{27\gamma-15\delta^2\gamma+2\Lambda R^2(3-5\delta^2)}{24\pi R^2} \,,
\end{equation}
\begin{equation}
\label{Pr2}
P_r(\delta) = \frac{3\gamma-3\delta^2\gamma+ 2\Lambda R^2(1-\delta^2)}{8\pi R^2} \,,
\end{equation} 
\begin{equation}
\label{Pt2}
P_\perp(\delta) = \frac{A_1(\delta)}{24\pi R^2 \left[ \Lambda R^2\delta^2(2\delta-3)+
3\delta^4\gamma-9\delta^2\gamma+3\right] } \,,
\end{equation} 
where $A_1(\delta)=8\Lambda^2R^4\delta^6 +3\Lambda R^2\left[\gamma\delta^2(8\delta^4-6\delta^2+9)+6(1-2\delta^2) \right] +9\gamma^2\delta^2(2\delta^4-3\delta^2+3) +27\gamma(1-2\delta^2) $.

In the next subsections we do  all  analyzes for $\Lambda R^2$ in order to avoid the arbitrary choice of a value for the radius of the object at that moment. The variable changing  considered here does not affect our conclusions since it does not modify the signs of the quantities examined, that are, pressures, density and inequalities imposed by the energy conditions. In the following we consider some particular values for $\Lambda R^2$ in order to clarify the analysis approach adopted.

\subsection{Analysis for \texorpdfstring{$\Lambda R^2=-0.1$}{Lg}}

\begin{figure}[H]
	 \subfloat[][\centering]
	{\includegraphics[width=8.0cm]{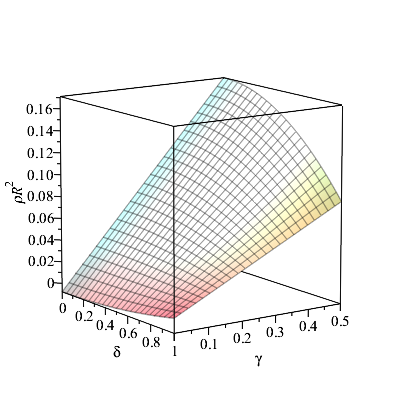}}
	\subfloat[][\centering]
    {\includegraphics[width=8.0cm]{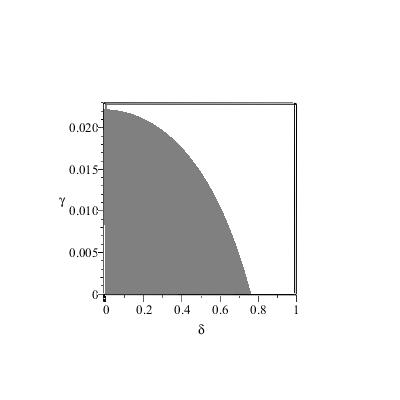}}
    \caption{In panel (a) we have $\rho R^2$ in terms of $\delta$ and $\gamma$, showing that $\rho R^2$ is negative for small values of $\gamma$ in some region inside the source. The gray region in panels (b) reveals the combinations of $\gamma$ and $\delta$ for which $\rho R^2<0$. \label{fig:E1}}
\end{figure}

\begin{figure}[H]
\centering
    \subfloat[][\centering]
    {\includegraphics[width=8.0cm]{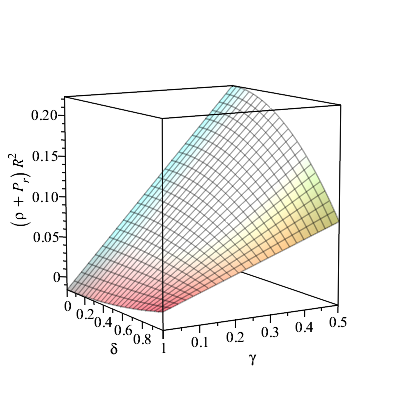}}
    \subfloat[][\centering]
    {\includegraphics[width=8.0cm]{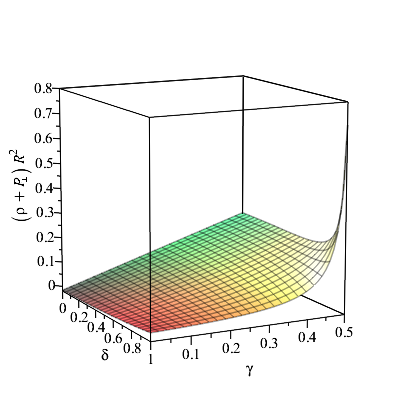}}\\
    \subfloat[][\centering]
    {\includegraphics[width=8.0cm]{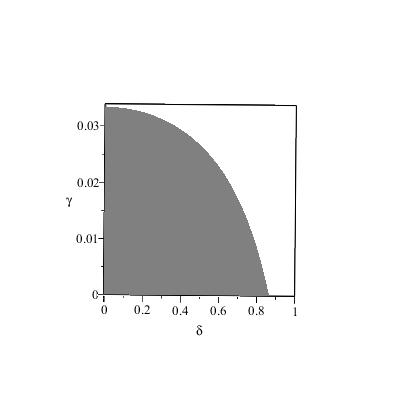}}
    \subfloat[][\centering]
    {\includegraphics[width=8.0cm]{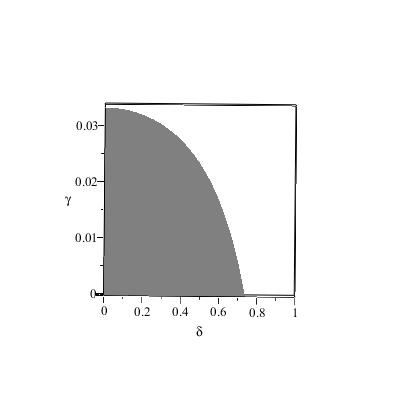}}
    \caption{Null energy conditions: {In panels (a) and (b) we have $(\rho+P_r)R^2$ and $(\rho+P_r)R^2$, respectively, in terms of $\delta$ and $\gamma$, showing that both energy conditions are negative for small values of $\gamma$ in some region inside the source. The gray regions in panel (c) and (d) reveal the combinations of $\gamma$ and $\delta$ for which each null energy condition defined on the top is violated. \label{fig:E2}}}
\end{figure}

\begin{figure}[H]
	\subfloat[][\label{fig:E3a}\centering]
	{\includegraphics[width=8.0cm]{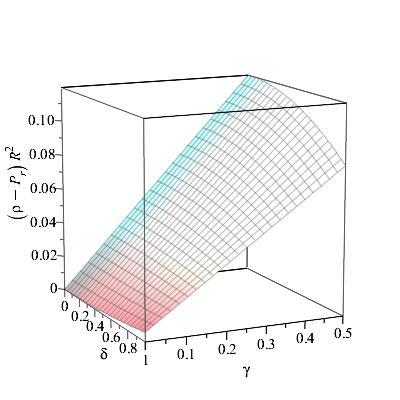}}
	\subfloat[][\label{fig:E3b}\centering]	
	{\includegraphics[width=8.0cm]{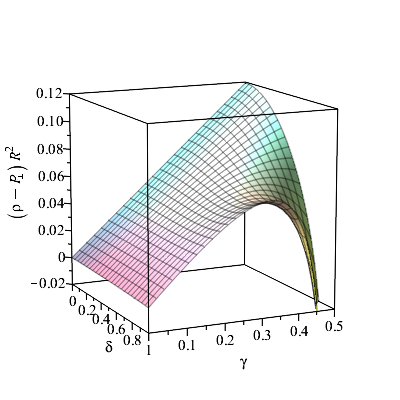}}\\
	\subfloat[][\centering]
    {\includegraphics[width=8.0cm]{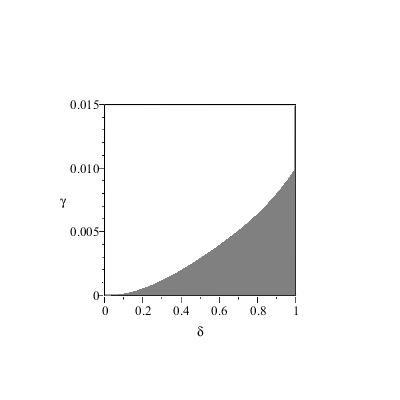}}
	\subfloat[][\centering]
    {\includegraphics[width=8.0cm]{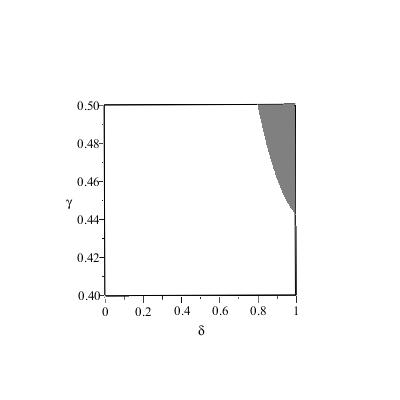}}
    \caption{In panels (a) and (b) we have $(\rho-P_r)R^2$ and $(\rho-P_r)R^2$, respectively, in terms of $\delta$ and $\gamma$. The gray regions in panel (c) and (d) show two combinations of $\gamma$ and $\delta$ for which $(\rho-P_\perp)R^2<0$. \label{fig:E3}}
\end{figure}

\begin{figure}[H]
	\subfloat[][\centering]
	{\includegraphics[width=8.0cm]{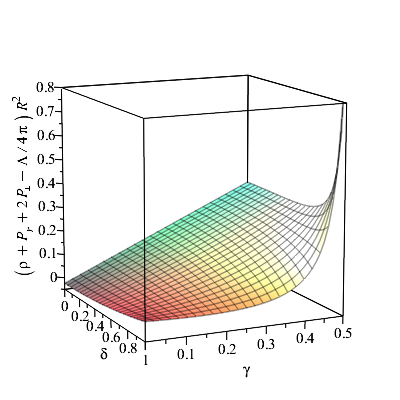}}
	\subfloat[][\centering]
    {\includegraphics[width=8.0cm]{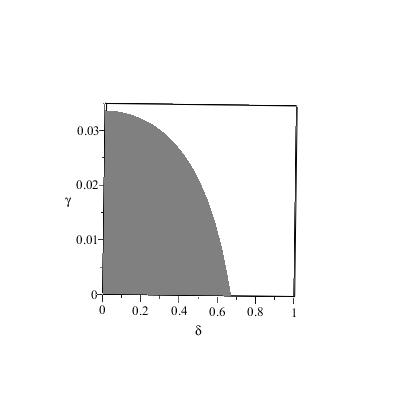}}
    \caption{The panel (a) presents one of the strong enrgy condition, $(\rho+P_r+2P_\perp-\Lambda/4\pi)R^2$, in terms of $\delta$ and $\gamma$, showing that it is violated for small values of $\gamma$ in some region inside the source. The gray region in panel (b) reveals the combinations of $\gamma$ and $\delta$ for which $(\rho+P_r+2P_\perp-\Lambda/4\pi)R^2<0$.\label{fig:E4}}
\end{figure}

From figure (\ref{fig:E1}), we see that there is a lower limit for the mass-radio ratio, that is $\gamma>0.0222$. By figures (\ref{fig:E2}a) and (\ref{fig:E2}b), we have $\gamma>0.0333$. In addition, from figure (\ref{fig:E3}b), we see that that $0.0101 < \gamma < 0.4420$. Lastly, figure (\ref{fig:E4}) shows us that $\gamma>0.0333$. Thus, the interval for $\gamma$ where all the energy conditions are satisfied is given by 
\begin{equation}
\label{intervalo1}
0.0333 < \gamma < 0.4420 \,.
\end{equation}

This model, addopting these specific values for $\Lambda$, related with the choice $\Lambda R^2=-0.1$, does not allow local dark energy inside the star, since there are no ranges for $\gamma$ in which only the strong energy condition is violated. This conclusion, however, can not be extended for other values of $\Lambda<0$.

\subsection{Analysis for \texorpdfstring{$\Lambda=0$}{Lg}}

\begin{figure}[H]
	\subfloat[][\label{fig:E11a}\centering]
	{\includegraphics[width=8.0cm]{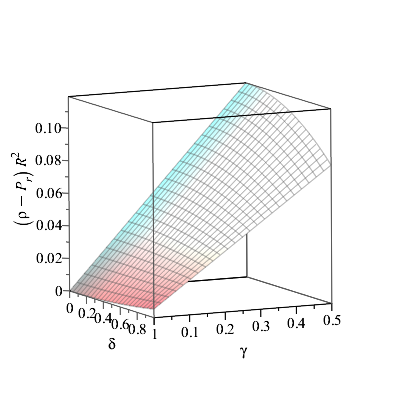}}
	\subfloat[][\label{fig:E11b}\centering]
	{\includegraphics[width=8.0cm]{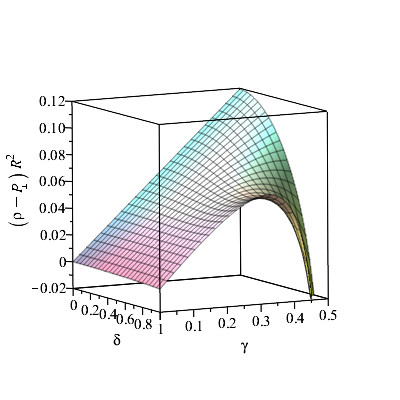}}\\
	\subfloat[][\centering]
	{\includegraphics[width=8.0cm]{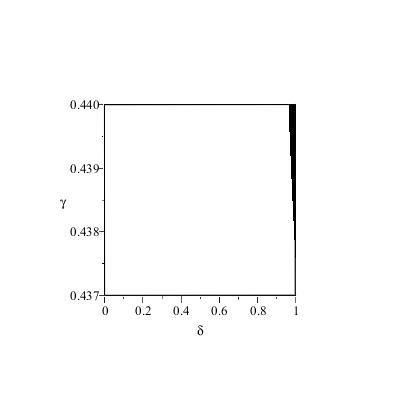}}
	\caption{In panels (a) and (b) we have $(\rho-P_r)R^2$ and $(\rho-P_r)R^2$, respectively, in terms of $\delta$ and $\gamma$. The black small region in panel (c) shows the combination of $\gamma$ and $\delta$ for which $(\rho-P_\perp)R^2<0$.\label{fig:E11}}
\end{figure}

\begin{figure}[H]
	\subfloat[][\centering]
	{\includegraphics[width=8.0cm]{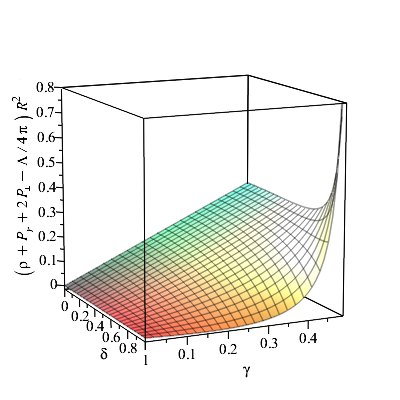}}
	\subfloat[][\centering]
	{\includegraphics[width=9.0cm]{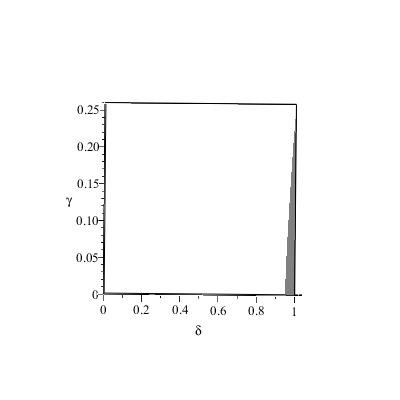}}
	\caption{The panel (a) presents one of the strong energy condition, $(\rho+P_r+2P_\perp-\Lambda/4\pi)R^2$, in terms of $\delta$ and $\gamma$, showing that it is violated for some values of $\gamma$ in the border of the source. The gray region in panel (b) reveals the combinations of $\gamma$ and $\delta$ for which $(\rho+P_r+2P_\perp-\Lambda/4\pi)R^2<0$.\label{fig:E12}}
\end{figure}

By the figure (\ref{fig:E11}b) we must imposes that $\gamma<0.4375$. On the other hand, from  (\ref{fig:E12}), we get that $\gamma>0.2500$. Therefore, we can express the interval where there is a violation of all energy conditions as

\begin{equation}
\label{intervalo5}
0.2500 < \gamma < 0.4375 \,.
\end{equation}

It is also possible to determine a range of $\gamma$ in which only the condition of strong energy is violated, featuring the presence of a local dark energy, that is,

\begin{equation}
\label{intervalo6}
0 < \gamma < 0.2500 \,.
\end{equation}  

In the figure (\ref{fig:E12}), we can see that the dark energy, when is present, appears in the outermost layers.

\begin{figure}[H]
	\centering	 
	\includegraphics[width=8.0cm]{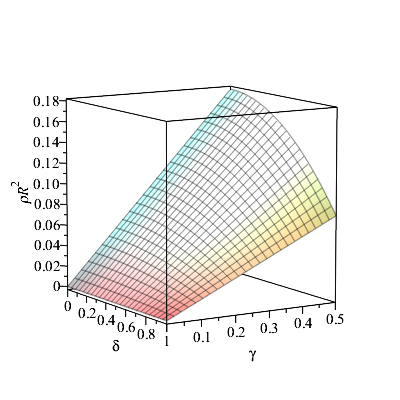}
		\caption{$\rho$\label{fig:E9}}
\end{figure}

\begin{figure}[H]
	\subfloat[][\label{fig:E10a}\centering]
	{\includegraphics[width=8.0cm]{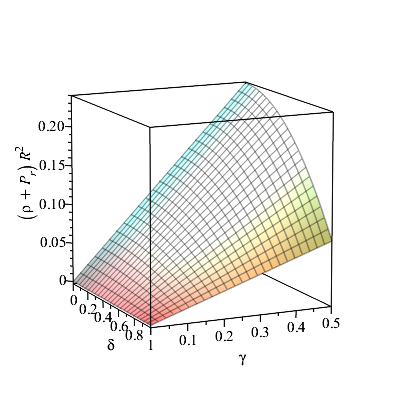}}
	\subfloat[][\label{fig:E10b}\centering]
	{\includegraphics[width=8.0cm]{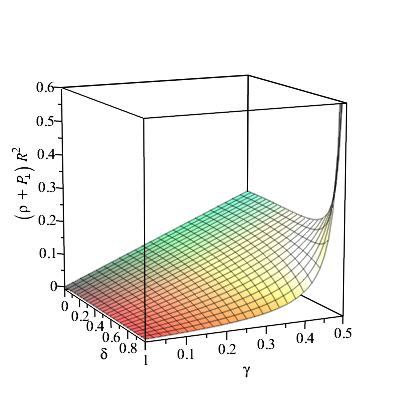}}
	\caption{$(\rho+P_r)R^2$ and $(\rho+P_\perp)R^2$\label{fig:E10}}
\end{figure}

\subsection{Analysis for \texorpdfstring{$\Lambda R^2=0.1$}{Lg}}

\begin{figure}[H]
		\subfloat[][\centering]
	{\includegraphics[width=8.0cm]{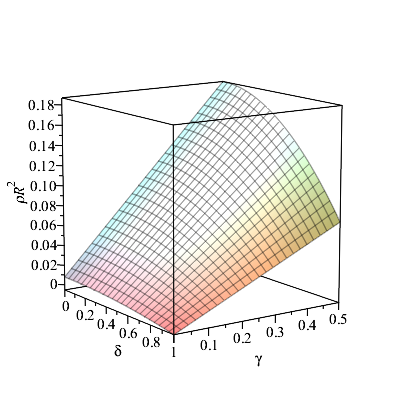}}
		\subfloat[][\centering]
	{\includegraphics[width=9.0cm]{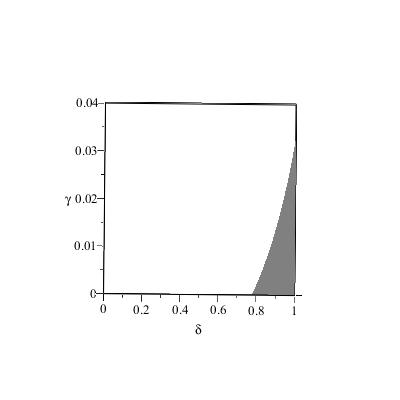}}
		\caption{In panel (a) we have $\rho R^2$ in terms of $\delta$ and $\gamma$, showing that $\rho R^2$ is negative for small values of $\gamma$ in some region inside the source. The gray region in panels (b) reveals the combinations of $\gamma$ and $\delta$ for which $\rho R^2<0$. \label{fig:E5}}
\end{figure}

\begin{figure}[H]
	\subfloat[][\centering]
	{\includegraphics[width=8.0cm]{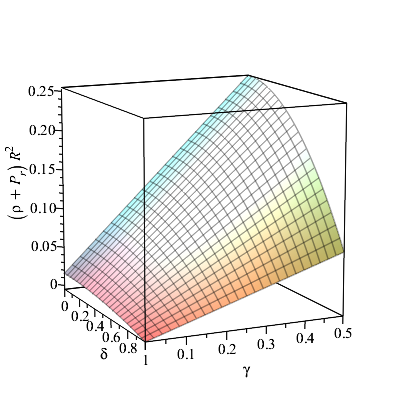}}
	\subfloat[][\centering]
	{\includegraphics[width=8.0cm]{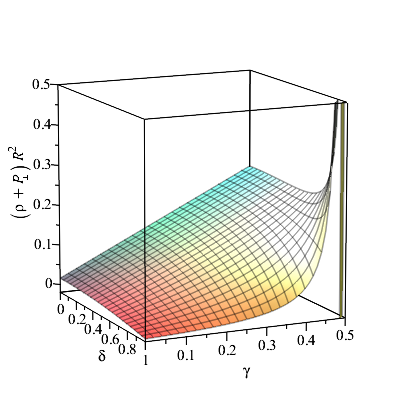}}\\
	\subfloat[][\centering]
	{\includegraphics[width=5.5cm]{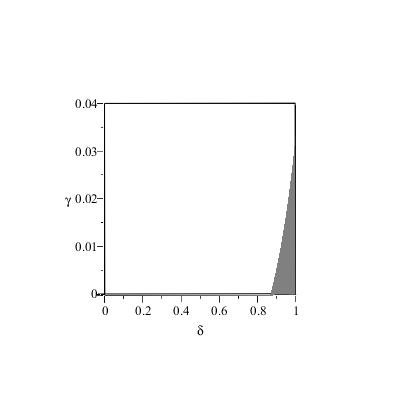}}
	\subfloat[][\centering]
	{\includegraphics[width=5.5cm]{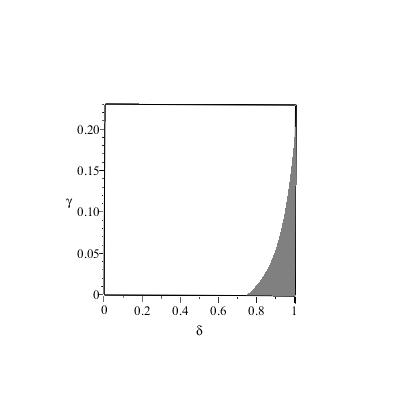}}
	\subfloat[][\centering]
	{\includegraphics[width=5.5cm]{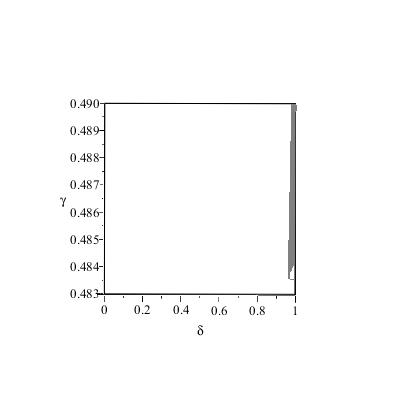}}
		\caption{Null energy conditions: In panels (a) and (b) we have $(\rho+P_r)R^2$ and $(\rho+P_\perp)R^2$, respectively, in terms of $\delta$ and $\gamma$. The gray regions in panel (c) show the combination of of $\gamma$ and $\delta$ for which $(\rho+P_r)R^2<0$ while in panels (d) and (e) appears the two set of combinations of $\gamma$ and $\delta$ for which $(\rho+P_\perp)R^2<0$. \label{fig:E6}}
\end{figure}

\begin{figure}[H]
	\subfloat[][\label{fig:E7a}\centering]
	{\includegraphics[width=8.0cm]{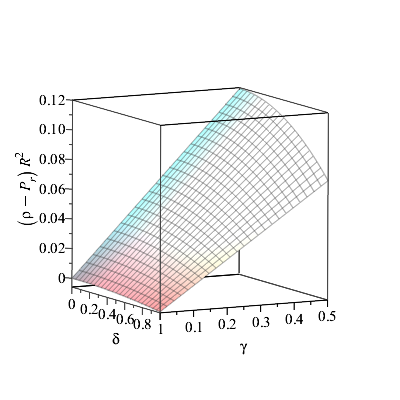}}
	\subfloat[][\label{fig:E7b}\centering]
	{\includegraphics[width=8.0cm]{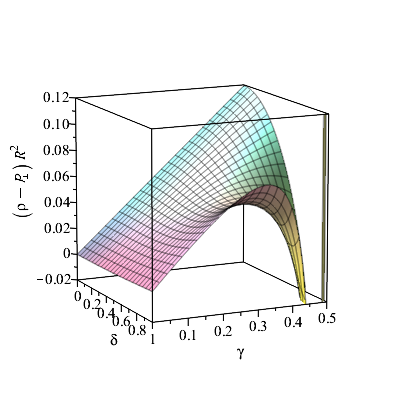}}\\
	\subfloat[][\centering]
	{\includegraphics[width=9.0cm]{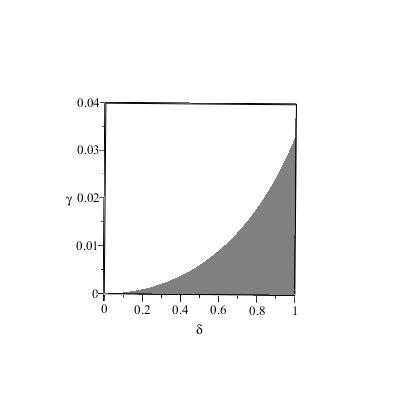}}
		\subfloat[][\centering]
	{\includegraphics[width=9.0cm]{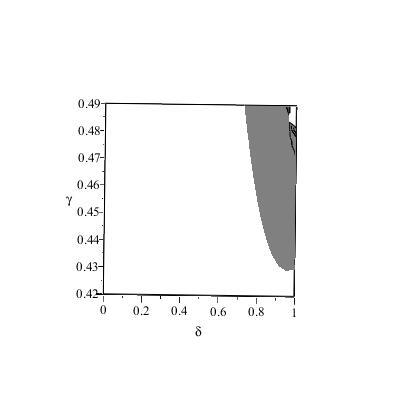}}
	\caption{In panels (a) and (b) we have $(\rho-P_r)R^2$ and $(\rho-P_\perp)R^2$, respectively, in terms of $\delta$ and $\gamma$, showing that both energy conditions are negative for small values of $\gamma$ in some region inside the source. The gray regions in panels (c) and (d) reveal the combinations of $\gamma$ and $\delta$ for which $(\rho-P_r)R^2<0$ and $(\rho-P_\perp)R^2<0$, respectively. \label{fig:E7}}
\end{figure}

\begin{figure}[H]
	\subfloat[][\centering]
	{\includegraphics[width=8.0cm]{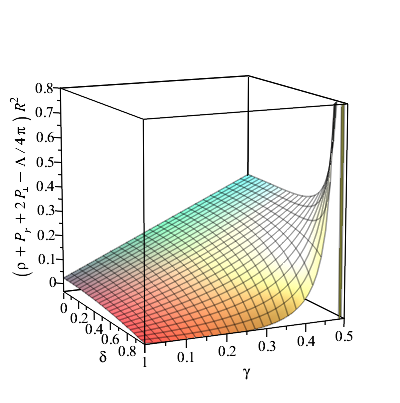}}
		\subfloat[][\centering]
	{\includegraphics[width=8.0cm]{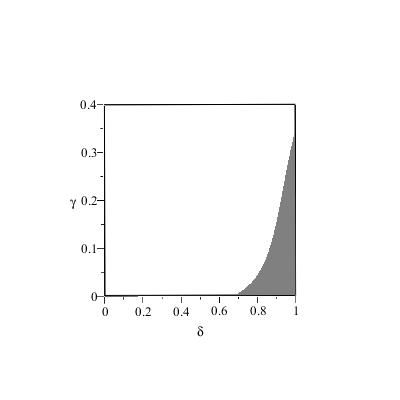}}\\
	\subfloat[][\centering]
	{\includegraphics[width=8.0cm]{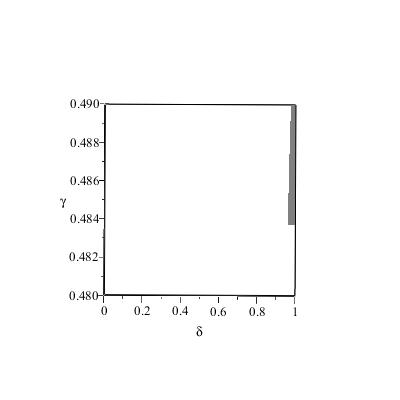}}
	\caption{The panel (a) presents one of the strong energy condition, $(\rho+P_r+2P_\perp-\Lambda/4\pi)R^2$, in terms of $\delta$ and $\gamma$, showing that it is violated for two different intervals of $\gamma$, but always near the border of the source. The gray regions in panels (b) and (c)  reveal the combinations of $\gamma$ and $\delta$ for which $(\rho+P_r+2P_\perp-\Lambda/4\pi)R^2<0$.\label{fig:E8}}
\end{figure}

Figures (\ref{fig:E5}), (\ref{fig:E6}a), and (\ref{fig:E7}a) show that $\gamma > 0.0333$. Figure (\ref{fig:E6}b) reveals that $0.2133 < \gamma < 0.4833$. From figure (\ref{fig:E7}b), we get $\gamma < 0.4320$. Finally, figure (\ref{fig:E8}) imposes $0.3461 < \gamma < 0.4833$. So, all the energy conditions are satisfied for the following interval
\begin{equation}
\label{intervalo2}
0.3461 < \gamma < 0.4320 \,.
\end{equation}

As figure (\ref{fig:E8}) is associated to the effective strong energy condition, we can express the interval where only it is violated as
\begin{equation}
\label{intervalo3}
0.2133 < \gamma < 0.3461 \,.
\end{equation}

Similarly to the case without a cosmological constant, here we have again that the effective dark energy, when is present, appears in the outermost layers. Note that the presence of this kind of effective dark energy inside the compact object was not induced exclusively by the cosmological constant, although it could be avoid if the last one is sufficiently negative, even having local dark energy inside the star.

In the previous subsections, we presented a detailed analysis only for the cases $\Lambda R^2=-0.1$, $\Lambda R^2=0$, and $\Lambda R^2=0.1$, but a more complete set of values of $\Lambda R^2$ is shown in the table below. There, we separated the range for $\gamma$ where all energy conditions are satisfied and that only the strong energy condition is violated. This last one means that there is a local dark energy trapped inside the star.

From table below, we can see that stars with local dark energy are favored for lower mass-radius ratios. For $\Lambda\geq 0$, the higher the $\Lambda$, the smaller the $\gamma$ range where we find models that satisfy all energy conditions or that admit effective dark energy in the stellar interior, and greater is the lower limit for $\gamma$. This could be interpreted as if the repulsion due to the positive $\Lambda$ demands a greater mass-radius ratio to balance the configuration. On the other hand, for $\Lambda<0$, the greater the absolute value of $\Lambda$, the greater the $\gamma$ interval that satisfies all energy conditions, while the $\gamma$ interval that allows this kind effective dark energy is narrowed, showing that negative $\Lambda$ disfavors the presence of this dark energy, as expected. In all cases, it is curious that this effective dark energy appears concentrated at the edge of the star. Considering the hypothesis, not yet completely confirmed, that the cosmological constant $\Lambda$ in fact is the responsible for the accelerated expansion of the universe, and taking its value from the cosmological observations ($\Lambda \approx 10^{-52} m^{-2}$), we can conclude that it does not interfere significantly in the structure of a compact object, at least for the model studied here. Taking the radius of a typical neutron star, between 10 and 15 kilometers, the smallest value of $\Lambda R^2$ that affects the $\gamma$ intervals in the table below, ie $\Lambda R^2=10^{-8}$, would correspond to $\Lambda=10^{-16}m^{-2}$. A similar conclusion was reached by \cite{BHP2016}, although they have considered a completely different model. In their work, it was obtained a equation of state through numerical computation for a many-body system. They concluded that  the influence of a positive cosmological constant, on the maximum mass for the neutron star, is considerable only for $\Lambda>10^{-14} m^{-2}$ and the maximum mass grows as $\Lambda$ increases.

\begin{table}
\begin{center}
	\begin{tabular}{ | c | c | c | }
		\hline
		$\Lambda R^{2}$ & All energy conditions are satisfied & Stars with effective/local dark energy \\ \hline
		$-10^{-1}$ & $0.03333333 < \gamma < 0.44202868$ & $\------------$ \\ \hline
		\multirow{2}{*}{$-10^{-2}$} & $0.00333333 < \gamma < 0.04091633$ & \multirow{2}{*}{\: $0.04091633< \gamma < 0.22491701$} \: \\ & $0.22491701 < \gamma < 0.43800070$ & \\ \hline
		\multirow{2}{*}{$-10^{-4}$} & $0.00003333 < \gamma < 0.00036699$ & \multirow{2}{*}{$0.00036699 < \gamma < 0.24979135$} \\ & $0.24979135 < \gamma < 0.43750658$ & \\ \hline
		$0$ & $0.25000000 < \gamma < 0.43750000$ & $0 < \gamma < 0.25000000$ $^*$ \\ \hline
		$10^{-8}$ & $0.25000002 < \gamma < 0.43750000$ & $0.00000003 < \gamma < 0.25000002$  \\ \hline
		$10^{-6}$ & $0.25000208 < \gamma < 0.43749995$ &  $0.00000333 < \gamma < 0.25000208$ \\ \hline
		$10^{-4}$ & $0.25020801 < \gamma < 0.43749494$ &  $0.00033315< \gamma < 0.25020801$ \\ \hline
		$10^{-2}$ & $0.26820932 < \gamma < 0.43698881$ &  $0.03161905 < \gamma < 0.26820932$ \\ \hline
		$10^{-1}$ & $0.34609364 < \gamma < 0.43192038$ &  $0.21333333 < \gamma < 0.34609364$ \\ 
		\hline
	\end{tabular}
\caption{Allowable compactness range. $^*$ Here, as $\Lambda = 0$, we have only local dark energy, that is, dark energy exclusively from the fluid, not effective dark energy.} 
\label{tab:tabela1}
\end{center}
\end{table}

\section{Other solutions without local dark energy}

\subsection{The Durgapal-Bannerji density profile}

This profile was proposed by Durgapal and Bannerji, here called as the Durgapal-Bannerji density profile, in order to construct an analytical isotropic compact object model \cite{DurgapalBannerji}. It is written as
\begin{equation}
\label{rho4}
\rho(r) = \frac{3C\left(3+Cr^2\right)}{16\pi(1+Cr^2)^2}\, ,
\end{equation}
with $C$ being an arbitrary constant to be determined.

It was considered in many later works, in different contexts, among which we can mention \cite{FinchSkea} and \cite{KomathirajMaharaj}.

Performing the same procedure as in the previous section, we determined
\begin{equation}
\label{h2}
h(r) = -\frac{\Lambda r^2}{3}-\frac{Cr^2-2}{2Cr^2+2} \,,
\end{equation}
\begin{equation}
\label{Pr4}
P_r(r) =\frac{\Lambda}{6\pi}- \frac{3C(Cr^2-1)}{16\pi(Cr^2+1)^2} \,,
\end{equation}
\begin{equation}
\label{Pt4}
P_\perp(r) = \frac{B_0(r)}{48\pi(Cr^2+1)^3 \left[3\left(Cr^2-2\right)+2\Lambda r^{2}(Cr^{2}+1)\right]} \,,
\end{equation}
where

\begin{eqnarray}
B_0(r)&=& 8\Lambda^{2}r^{2}(Cr^2+1)^4 +6\Lambda\left(4C^4r^8-17C^3r^6-42C^2r^4-29Cr^2-8\right) \nonumber\\
&-&27C\left(3C^2r^4-Cr^2+2\right) \,.\nonumber
\end{eqnarray}

We have isotropic pressures in the center of the star, and taking $P_r$ on the hypersurface and applying the conditions given by (\ref{PrR0}), we find

\begin{equation}
\label{Lambda1}
\Lambda = \frac{9C(CR^2-1)}{8(CR^2+1)^2} \,.
\end{equation}
where $0\le r\le R$.

Thus, substituting (\ref{Lambda1})
in (\ref{h2}), and taking $r=R$, we have
\begin{equation}
\label{hR3}
h^{-}(R)=\frac{8-7CR^2(CR^2+1)}{8(CR^2+1)^2} \,.
\end{equation}

On the other hand, considering the exterior metric, we get
\begin{equation}
\label{hR4}
h^{+}(R)=1-2\gamma+\frac{3CR^2(CR^2-1)}{8(CR^2+1)^2} \,.
\end{equation}

Imposing the continuity of the metric $h^{-}=h^{+}$ on the hypersurface $\Sigma$, it is possible to determine the parameter $C$, that is,
\begin{equation}
\label{C1}
CR^{2} = \frac{4\gamma}{(3-4\gamma)} \,.
\end{equation}

Since $\rho$ must be positive for any value of $r$, we must impose that $C>0$, which implies
\begin{equation}
\label{gama1}
0 \leq \gamma < 0.750 \,.
\end{equation}

In order to obtain the cosmological constant in terms of the compactness, we have to use (\ref{C1}) in (\ref{Lambda1}). So,
\begin{equation}
\label{Lambda20}
\Lambda R^2 = \frac{(8\gamma-3)\gamma}{2} \,.
\end{equation}

From the last equation, we can see that $\Lambda = 0$ for $\gamma=0$, showing that this model does not admit the de Sitter spacetime as a limit, and for $\gamma=0.375$. For $0\leq\gamma<0.375$, $\Lambda$ must be negative, while for $0.375<\gamma<0.750$, $\Lambda$ must be positive.

Substituting (\ref{C1}) and (\ref{Lambda20}) into (\ref{h2}), we get
\begin{equation}
\label{h3}
h(\delta) = \frac{18+32\gamma^3\delta^2(1-\delta^2)-12\gamma^2\delta^2(3-\delta^2)-3\gamma(8+\delta^2)}{18-24\gamma(1-\delta^2)} \,,
\end{equation}
with $\delta=r/R$.

\begin{figure}[H]
\subfloat[][\centering]
{\includegraphics[width=8.0cm]{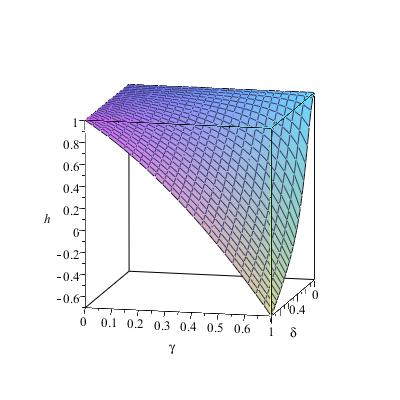}}
\subfloat[][\centering]
{\includegraphics[width=9.0cm]{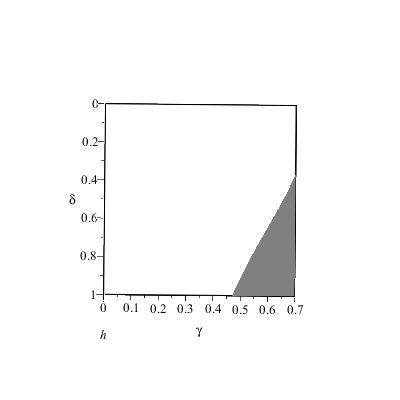}}
\caption{In panel (a), we have $h$ in terms of $\delta$ and $\gamma$. We can see that $h$ is negative for some values of $\gamma$. The gray region in panels (b) reveals the combinations of $\gamma$ and $\delta$ for which $h<0$. \label{fig:E21}}
\end{figure}

Imposing that $h(\delta)$ must be positive anywhere in the interior spacetime (\ref{h3}), we have
\begin{equation}
\label{gama2}
0 \leq \gamma \leq 0.470 \,.
\end{equation}

\begin{figure}[H]
\subfloat[][\centering]
{\includegraphics[width=8.0cm]{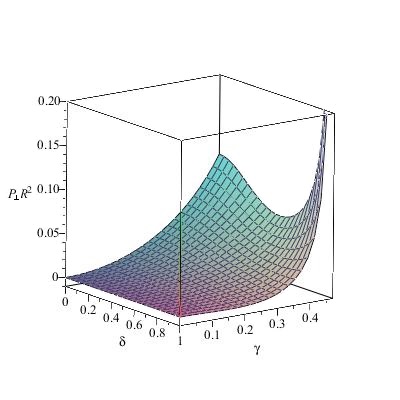}}
\subfloat[][\centering]
{\includegraphics[width=9.0cm]{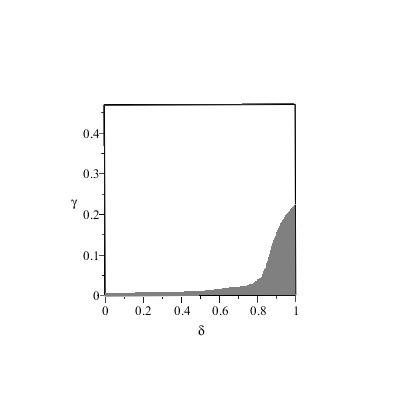}}
\caption{In panel (a), we have $P_\perp R^2$ in terms of $\delta$ and $\gamma$, showing that $P_\perp$ is negative for some values of $\gamma$. The gray region in panels (b) reveals the combinations of $\gamma$ and $\delta$ for which $P_\perp<0$. \label{fig:E19}}
\end{figure}

\begin{figure}[H]
\subfloat[][\centering]
{\includegraphics[width=8.0cm]{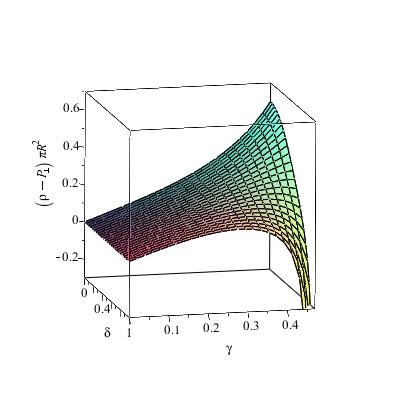}}
\subfloat[][\centering]
{\includegraphics[width=9.0cm]{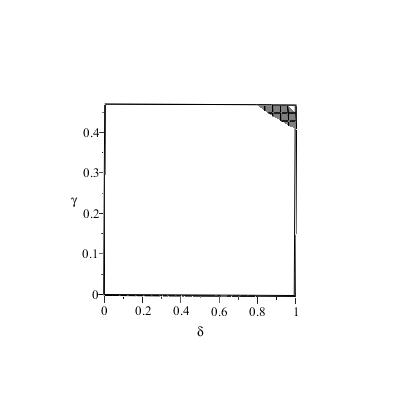}}
\caption{In panel (a), we have $(\rho -P_\perp)R^2$ in terms of $\delta$ and $\gamma$. Note that $\rho -P_\perp$ is negative for values of $\gamma$ near its upper limit ($\approx 0.411$), and in the border of the object. The gray region in panels (b) reveals the combinations of $\gamma$ and $\delta$ for which $\rho -P_\perp<0$. \label{fig:E20}}
\end{figure}

Now let us analyse the behaviour of the fluid in terms of pressures and energy conditions. We verified that the energy density and radial pressure are positive throughout the source, but that the tangential pressure is negative in some regions, for models with compactness in the interval $0\le\gamma <0.223$, as shown in figure (\ref{fig:E19}). Note that this range of $\gamma$ requires a negative cosmological constant. For models with $\gamma <0.411$, we find that all energy conditions are satisfied, including the local or effective strong energy condition. This restriction for $\gamma$ is imposed by the dominant energy condition $\rho-P_\perp>0$, which can be seen in figure (\ref{fig:E20}). So, if we consider a universe with positive cosmological constant, as pointed out by the observations, this model admits objects with compactness values in the interval $0.375<\gamma<0.411$, without any possibility of including local or effective dark energy.

\subsection{Null radial pressure} 

Our propose is to verify if it is possible a model without radial pressure of the matter, sustained only by a tangential pressure, with the cosmological constant.

Imposing that the junction, energy and regularity conditions are satisfied, we find
\begin{equation}
\label{Pr(0)h4}
h(r) = 1 - \frac{3\gamma\delta^2}{2R^2} \,,
\end{equation}
\begin{equation}
\label{Pr(0)rho3}
\rho(r) = \frac{3\gamma}{4\pi R^2} \,,
\end{equation}
\begin{equation}
\label{Pr(0)Pt5}
P_\perp(r) = \frac{9\gamma^2 \delta^2}{8\pi R^2(2-3\gamma\delta^2)} \,,
\end{equation}
with
\begin{equation}
\label{Pr(0)Lambda1}
\Lambda = -\frac{3\gamma}{2R^2} \,,
\end{equation}
where $\gamma = M/R$ and $\delta=r/R$. From (\ref{Pr(0)Pt5}), we can see that the isotropy in the pressures is assure in the center, as expected.

It is easy to see that all the energy conditions are simultaneously satisfied, for every radius inside the source, in the limit
\begin{equation}
\label{Casopart16}
0 \le \gamma \le \frac{4}{9}\,,
\end{equation}
coinciding with the Buchdahl limit \cite{Buchdahl}, However, in that case, there is also a constant energy density, but pressures are isotropic and upper limit of $\gamma$ corresponds to an infinity pressure, which ensures the stability of the star. Here, the superior limit for $\gamma$ in (\ref{Casopart16}) is imposed by the dominant energy condition ($\rho-P_\perp \ge 0$).

\section{Conclusion}

Here we start from a static spherically symmetric matter distribution with anisotropic pressure, in the context of the modified Einstein's field equation by the inclusion of a cosmological constant. Then, we can see that the differential equations imply directly a non-local equation of state for the fluid, already pointed out by Hernández and Núñez \cite {HN2004}, but now with the cosmological constant. We considered all usual physical restrictions, such as the hydrostatic equilibrium, regularity, junction, and energy conditions. In the sequence, we applied the results for two particular density profiles, which are called Stewart's profile \cite{Stewart} and Durgapal-Bannerji profile \cite{DurgapalBannerji}. We also considered the possibility of building models for which the radial pressure is zero. 

The model with the Stewart's profile proved to be particularly interesting, since, in addition to satisfying all the energy conditions for a certain range of the mass-radius ratio, it also admits another range of mass-radius ratio where only the strong energy condition is violated in the outermost layers, characterizing the presence of local, or effective, dark energy in its constitution. In a general way, we can say that stars with dark energy are favored for lower mass-radius ratios. The dependence of these ranges with the cosmological constant is discussed, as following: for $\Lambda\geq 0$, the higher the $\Lambda$, the smaller the $\gamma$ range where we find models that satisfy all energy conditions or that admit local, or effective, dark energy in the stellar interior, and greater is the lower limit for $\gamma$. This could be interpreted as if the repulsion due to the positive $\Lambda$ demands a greater mass-radius ratio to balance the configuration. For $\Lambda<0$, the greater the absolute value of $\Lambda$, the greater the $\gamma$ interval that satisfies all energy conditions, while the $\gamma$ interval that allows local, or effective, dark energy is narrowed, showing that negative $\Lambda$ disfavors the presence of this kind of dark energy. In all cases, it is curious that the dark energy appears concentrated at the edge of the object.

The configuration constructed from the Durgapal and Bannerji density profile can satisfy all the energy conditions for $0\leq\gamma\leq 0.411$, although for $0<\gamma< 0.375$ the model requires $\Lambda<0$, while for $0.375<\gamma< 0.411$, $\Lambda$ must be positive. Unlike the previous model, it does not admit dark energy in the composition of the object.

The model with $P_r=0$, the cosmological constant must be negative, and it is noteworthy that this model is able to describe a static configuration, in which the tangential pressure combined with a negative cosmological constant keeps the system in balance.

Here we consider three particular models, the first one that uses the energy density profile originally proposed by Stewart, the second one consider the energy density profile proposed by Durgapal and Bannerji and a third one, for which we did not adopt any density profile a priori, but imposed $P_r=0$. All of these have anisotropy in pressure and do not admit isotropy as a limit. Although we do not know a general demonstration, it seems reasonable to us that the dark energy in such objects, resulting from a sufficient negativity of pressure, can have its stability compromised if  this occurs in the same way in all directions. If the pressure is anisotropic, each direction could contribute in a different way, allowing more possibilities to produce stable configurations. For gravastar models, for example, which the TOV equation is considered for static solutions with negative pressures in the center, it was showed that anisotropic pressure is necessary \cite{CattoenEtAl}.

\section*{Acknowledgments}

LSMV and AB thank financial assistance from Coordenação de Aperfeiçoamento de Pessoal de Nível Superior (CAPES) -- Brazil and Fundação Carlos Chagas Filho de Amparo à Pesquisa do Estado do Rio de Janeiro (FAPERJ) -- Brazil. The author (MFAdaS) acknowledges the financial support from Financiadora de Estudos e Projetos -- FINEP -- Brazil, Fundação de Amparo à Pesquisa do Estado do Rio de Janeiro -- FAPERJ -- Brazil and Conselho Nacional de Desenvolvimento Cient\'{\i}fico e Tecnol\'ogico -- CNPq -- Brazil.

\end{document}